\documentclass[preprint,12pt]{elsarticle}
\usepackage{graphicx}

\journal{AAP. Published as Astropart.Phys. 36 (2012) 37-41
[arXiv:1009.1675]}

\begin{document}
\begin{frontmatter}

\title{A theoretical diagnosis on light speed anisotropy from GRAAL experiment}
\author[1]{Zhou Lingli} \ead{zhoull@pku.edu.cn}
\author[1,2,3]{Bo-Qiang Ma}
\ead{mabq@pku.edu.cn}
\address[1]{School of Physics
and State Key Laboratory of Nuclear Physics and Technology, \\Peking
University, Beijing 100871, China}
\address[2]{Center for High Energy
Physics, Peking University, Beijing 100871, China}
\address[3]{Center for History and Philosophy of Science, Peking University, Beijing 100871,
China}

\begin{abstract}
The light speed anisotropy, i.e., the variation of the light speed
with respect to the direction in an ``absolute" reference frame, is
a profound issue in physics. The one-way experiment, performed at
the GRAAL facility of the European Synchrotron Radiation Facility
(ESRF) in Grenoble, reported results on the light speed anisotropy
by Compton scattering of laser photons on high-energy electrons. So
far, most articles concerned with the GRAAL data have established
only the upper bounds on the anisotropy parameters based on
available theories. We use a new theory of the Lorentz invariance
violation to analyse the available GRAAL data and obtain the
stringent upper limit of the order $2.4\times10^{-14}$ on the
Lorentz violation parameters. In the meantime, we also can reproduce
the allowed light speed anisotropy appearing in the azimuthal
distribution of the GRAAL experimental data, and find that the
best-fit parameters are compatible with the competitive upper
bounds.
\end{abstract}
\end{frontmatter}
\newpage

Isotropy and constancy of light speed are two basic properties of
light in modern physics. Any evidence for their variation, even very
tiny, will have profound significance in science. The anisotropy of
the light speed in vacuum has been studied for more than 100 years,
and the most famous exploration is the Michelson-Morley experiment
in 1887~\cite{Michelson87}. There are many modern analogies of
experiments for the same purpose and most of them adopted round-trip
or two-way path propagation of light involving averaged light speed.
Therefore one-way experiments, which are sensitive to the first
order of light speed variation, deserve particular attention. In
this paper we provide a theoretical analysis of the results from the
one-way
experiment~\cite{Gurzadyan05,Gurzadyan07,Gurzadyan10,Bocquet10}
performed at the GRAAL facility of the European Synchrotron
Radiation Facility (ESRF) in Grenoble.

Most papers concerned with the GRAAL data just reported the upper
limits on the Lorentz violation parameters measuring the light speed
anisotropy by the commonly used theory. We use a new
theory~\cite{Ma10,SMS3,photon2} of the Lorentz invariance Violation
(LV) to analyse the GRAAL data. The Lorentz violation is related
with the space-time anisotropy, and then the space-time anisotropy
is the source for the light speed anisotropy. Therefore one can
understand the anisotropy of the propagation velocity of free
photons from the Lorentz violation, and on the other hand the
constraints on the light speed anisotropy put bounds on the Lorentz
violation parameters of the theory. In the new theory, the Lorentz
violation information is measured by a new matrix, denoted as
Lorentz invariance violation matrix. With the GRAAL data, we can
establish the conservative bounds on the elements of the Lorentz
violation matrix of free photons. At the same time, when the GRAAL
data are best-fitted, we can obtain the values of the Lorentz
violation parameters to observe possible anomaly implied by the data
already.


In the following, we prepare the knowledge needed first, including
reviewing the experiments and presenting the new model for photons.
Then we can get the upper constraints on the violation parameters
and also their best-fit values by analyzing the GRAAL experimental
data.

We provide a brief review on the principle~\cite{Gurzadyan96} of the
GRAAL
experiment~\cite{Gurzadyan05,Gurzadyan07,Gurzadyan10,Bocquet10}, in
which the highly monochromatic electrons are scattered on the laser
photons, for the study of light speed anisotropy in the ``absolute"
inertial frame at rest defined by null dipole of the Cosmic
Microwave Background (CMB) radiation. In the head-on Compton
scattering of the ultra-energy electrons and the low energy photons, the
energy $E$ of the scattered photon is given by
\begin{equation}\label{photonSE1}
E=\frac{4\gamma^2E_0}{1+4\gamma E_0/m_e+\theta^2\gamma^2},
\end{equation}
where $\theta$ is the angle between the scattered photon and the
incident electron, $E_0$ is the energy of the incident photon, and
$m_e$ and $\gamma$ are the mass and Lorentz factor of the incident
electron. On the other hand, the scattered electrons will separate
from the main incident electrons beam, and there is a distance $x$
between these two trajectories.
The energy of the scattered photon can also be written as
\begin{equation}\label{photonSE2}
E=\frac{E_e x}{A+x},
\end{equation}
with $E_e$ being the energy of the incident electron and $A$ being a
constant related with the experiment set-up. The maximum energy $E$ of the Compton scattered photons is
called as the Compton Edge (CE). From Eq.(\ref{photonSE1}), CE can
be obtained for $\theta=0$. In this case, using
Eqs.~(\ref{photonSE1}) and (\ref{photonSE2}), we get
\begin{equation}
x_{\mathrm{CE}}=\frac{4A\gamma E_0}{m_e}.
\end{equation}
So
\begin{equation}\label{xCE0}
\delta
x_{\mathrm{CE}}=\frac{4AE_0}{m_e}\delta\gamma=-\frac{4AE_0}{m_e}\beta^2\gamma^3\delta
c,
\end{equation}
where $\gamma=(1-\beta^2)^{-1/2}$ is the Lorentz factor of the
incident electron. At the GRAAL facility, the mean energy $E_e$ of
the electron beam is 6.04~GeV, i.e., $\gamma=11820$, three UV laser
lines around 351~nm and a green line at 512~nm are used, and
$A=159.28 \pm 0.2$~mm.

\begin{figure}
\centering
\includegraphics[width=.8\textwidth]{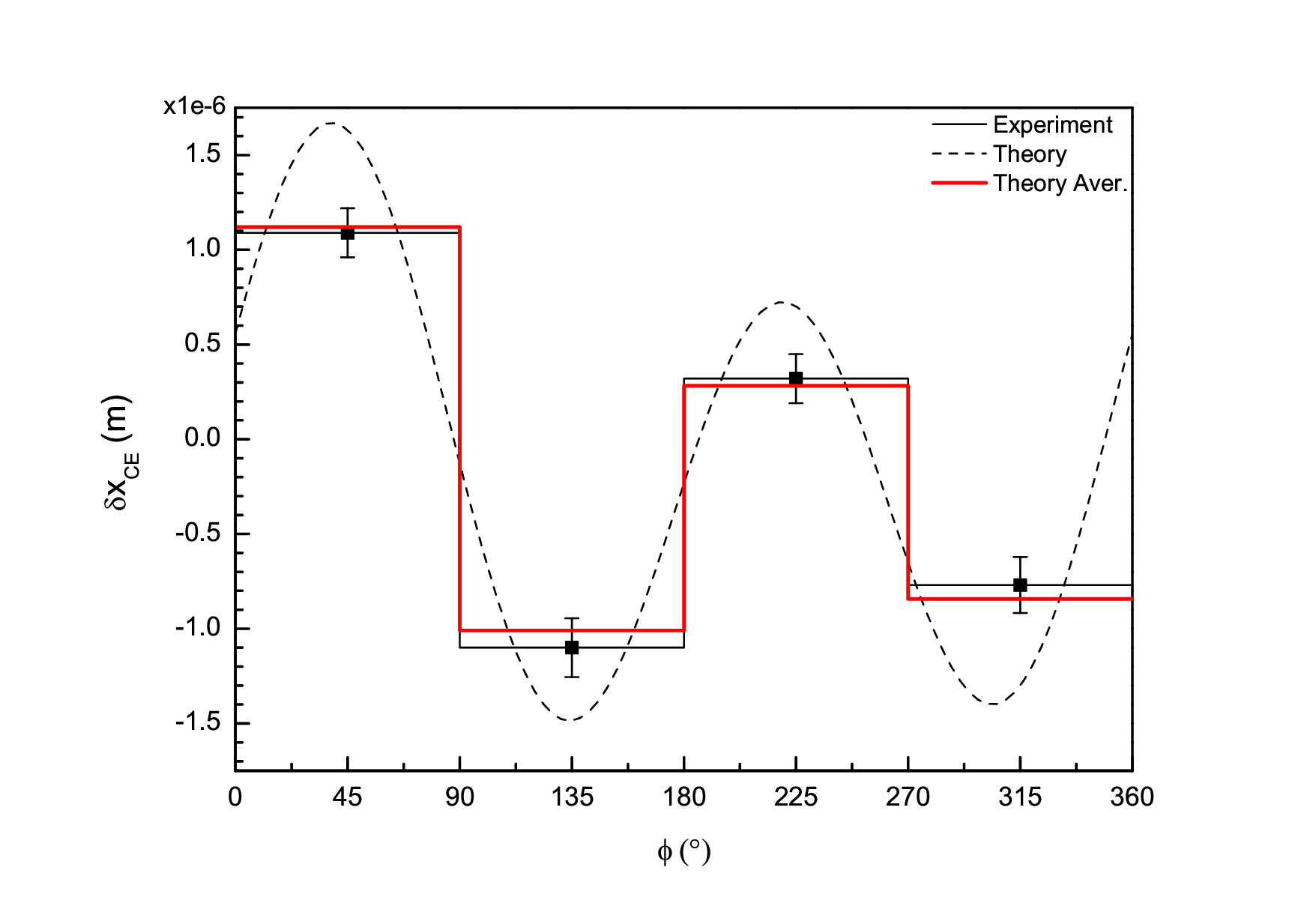}
\caption{$\delta x_{\mathrm{CE}}$ azimuthal distribution vs angles
of the GRAAL data of the years 1998-2005 on a plane ($x$-$y$ plane
or $\theta=\pi/2$). $\xi=-2.89\times10^{-13}$,
$\lambda=6.53\times10^{-14}$.}\label{fig1}
\end{figure}

\begin{figure}
\centering
\includegraphics[width=.8\textwidth]{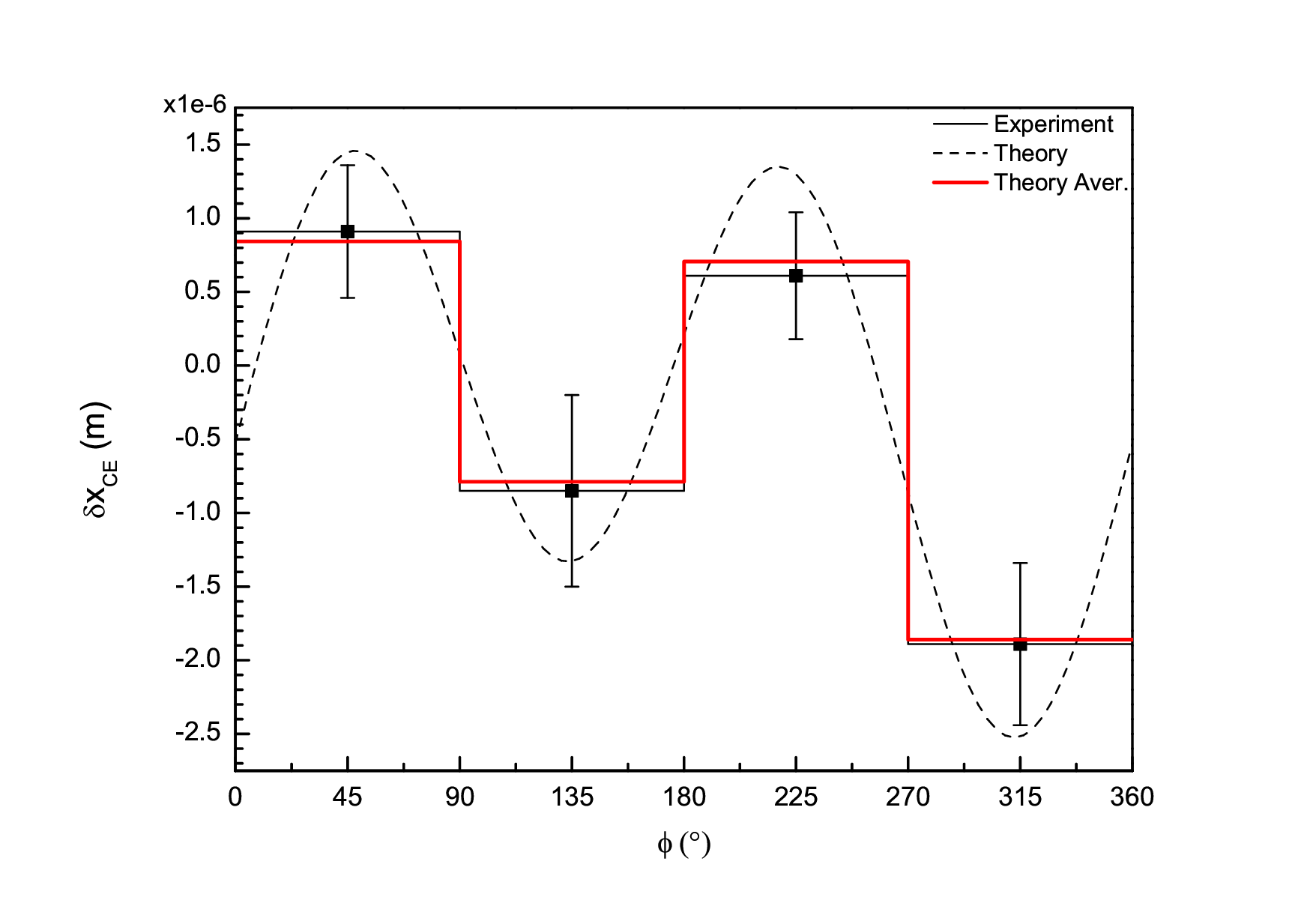}
\caption{$\delta x_{\mathrm{CE}}$ azimuthal distribution vs angles
of the GRAAL data of the year 2008 on a plane ($x$-$y$ plane or
$\theta=\pi/2$). $\xi=-3.64\times10^{-13}$,
$\lambda=8.24\times10^{-14}$.} \label{fig2}
\end{figure}

Now, $x_{\mathrm{CE}}$ has been measured in the GRAAL
experiment~\cite{Gurzadyan05,Gurzadyan07,Gurzadyan10,Bocquet10}.
With the given stable energies $E_0$ and $E_e$, $x_{\mathrm{CE}}$
varies over different directions in the space when $\delta c$ is
azimuthal dependent. The azimuthal distribution for
$x_{\mathrm{CE}}$ measured in the experiment determines the light
speed anisotropy $\delta c$. The GRAAL results for the measured
distance $\delta x_{\mathrm{CE}}$ related with the CE are shown in
Fig.~\ref{fig1} for the data of the years
1998-2005~\cite{Gurzadyan07} and in Fig.~\ref{fig2} for the data in
the year 2008~\cite{Gurzadyan10}, revealing the robust CE azimuthal
variation. The limits on the light speed anisotropy are reported in
Ref.~\cite{Bocquet10}, in which the azimuthal distribution presented
in Figs.~\ref{fig1}, \ref{fig2} was not discussed. We show in this
paper that the GRAAL results in Figs.~\ref{fig1}, \ref{fig2} can be
elegantly reproduced by our new theory of Lorentz
violation~\cite{Ma10,SMS3,photon2}, briefly illustrated below for
the free photon sector.

Nowadays, Lorentz invariance violation has triggered more and more
interests in physics (see, e.g.,
Refs.~\cite{Amelino06,Amelino08,Mattingly:2005re,Sidharth08,Xiao09,Xiao09b,Shao10,Shao10b,Shao10c}
and references therein). Especially the variation of the light speed
due to the effect of quantum space-time has received particular
attentions together with phenomenological
supports~\cite{AmelinoCamelia:1997gz,Gambini:1998it,AmelinoCamelia:2000mn,Myers:2003fd}.
Our framework~\cite{Ma10,SMS3} is a new fundamental theory of
Lorentz invariance violation from basic principles instead of from
phenomenological considerations. We proposed a general principle of
physical independence of the mathematical background manifold, i.e.,
the equations describing the laws of physics have the same form in
all admissible mathematical manifolds. Based on such principle, we
revealed the replacements $\partial^{\alpha} \rightarrow
M^{\alpha\beta}\partial_{\beta}$ and $D^{\alpha}\rightarrow
M^{\alpha\beta}D_{\beta}$ for the ordinary partial
$\partial_{\alpha}$ and the covariant derivative $D_{\alpha}$.
$M^{\alpha\beta}$ is a local matrix called Background Matrix (BM)
and can be divided into the sum of two matrices, i.e., $M^{\alpha
\beta}=g^{\alpha \beta}+\Delta^{\alpha \beta}$, where
$g^{\alpha\beta}$ is the metric of space-time and $\Delta^{\alpha
\beta}$ is a new matrix which brings new terms violating Lorentz
invariance in the standard model, therefore we denote the new
framework as the Standard Model Supplement (SMS). Then the
Lagrangian for the free gauge particle photon reads
\begin{eqnarray}
\mathcal{L}_{\mathrm{G}}&=&-\frac{1}{4}F^{\alpha\beta }F_{\alpha\beta}
-F_{\mu\nu}\Delta^{\mu\alpha}\partial_{\alpha}A^{\nu}\label{Lagrangian} \nonumber\\
&-&\frac{1}{2}\Delta^{\alpha\beta}\Delta^{\mu\nu}(g_{\alpha\mu}\partial_{\beta}
A^{\rho}\partial_{\nu}A_{\rho}-\partial_{\beta}A_{\mu}
\partial_{\nu}A_{\alpha}).
\end{eqnarray}
Since $\Delta^{\alpha\beta}$ contains all the LV information for the
space-time, we call it Lorentz invariance Violation Matrix (LVM).
All the LV effects vanish when $\Delta^{\alpha \beta}=0$. More
details of this new framework and its connection with available
phenomenological constraints on LV effect of photons can be found in
Refs.~\cite{Ma10,SMS3,photon2}.

We thus get the modified Maxwell equation (or motion equation)
\begin{equation}\label{maxwell1}
\Pi^{\gamma\rho}A_{\rho}=0,
\end{equation}
where $\Pi^{\gamma\rho}$ is also the inverse of the photon propagator
\begin{eqnarray}
\Pi^{\gamma\rho}
&=&-g^{\gamma\rho}\partial^2+\partial^{\gamma}\partial^{\rho}\nonumber\\
&+&\Delta^{\gamma\alpha}\partial^{\rho}\partial_{\alpha}+\Delta^{\rho\alpha}\partial^{\gamma}\partial_{\alpha}
+\Delta^{\gamma\beta}\Delta^{\rho\nu}\partial_{\beta}\partial_{\nu}\nonumber\\
&-&g^{\gamma\rho}(2\Delta^{\mu\alpha}\partial_{\mu}\partial_{\alpha}
+g_{\alpha\mu}\Delta^{\alpha\beta}\Delta^{\mu\nu}\partial_{\beta}\partial_{\nu}).
\end{eqnarray}
Three terms $\partial^{\gamma}\partial^{\rho}$,
$\Delta^{\gamma\alpha}\partial^{\rho}\partial_{\alpha}$ and
$\Delta^{\rho\alpha}\partial^{\gamma}\partial_{\alpha}$ (symmetric
for the indices $\gamma$ and $\rho$) can be omitted under the
consideration of the Lorentz gauge condition
$\partial^{\alpha}A_{\alpha}=0$  for the gauge field. With the
Fourier decomposition $A_{\rho}=\int dp A_{\rho}(p)e^{-ip\cdot x}$,
we can re-write Eq.~(\ref{maxwell1}) as
$$\Pi^{\gamma\rho}(p)A_{\rho}(p)=0,$$ where
\begin{eqnarray}
\Pi^{\gamma\rho}(p)&=&g^{\gamma\rho}(p^2+g_{\alpha\mu}\Delta^{\alpha\beta}\Delta^{\mu\nu}p_{\beta}p_{\nu}+2\Delta^{\alpha\beta}p_{\alpha}p_{\beta})\nonumber\\
&-&\Delta^{\gamma\beta}\Delta^{\rho\nu}p_{\beta}p_{\nu},\nonumber
\end{eqnarray}
which is the inverse of the free photon propagator in the momentum
space. A general parameterization for $p_{\alpha}$ can be done with
spherical coordinates, so $p_{\alpha}$ can be expressed as $(E$,
$-|\vec{p}|\sin{\theta}\cos{\phi}$,
$-|\vec{p}|\sin{\theta}\sin{\phi}$, $-|\vec{p}|\cos{\theta})$, where
the light speed constant is $c=1$. We find that there is a zero
eigenvalue and a corresponding eigenvector $A_{\rho}(p)$ for the
matrix $\Pi^{\gamma\rho}(p)$. So the determinant must be zero for
the existence of the solution $A_{\rho}(p)$
\begin{equation}\label{det}
\mathrm{det}(\Pi^{\gamma\rho}(p))=0.
\end{equation}
Then we have the equation
$$\sum_{i=0}^{8}\lambda_{i}(\Delta^{\alpha\beta},\theta,\phi)E^{i}|\vec{p}|^{8-i}=0.$$
The coefficient $\lambda_{i}(\Delta^{\alpha\beta},\theta,\phi)$ is a
variable related to the LVM $\Delta^{\alpha\beta}$ and the angles
$\theta$ and $\phi$. So there are 8 real solutions  for
$E(|\vec{p}|)$ at most, and in general there are no analytical
solutions for a general high order linear equation. But for some
simple cases of the LVM $\Delta^{\alpha\beta}$, we expect some
analytical solutions for $E$. Anyway, $E$  can be solved formally as
$E=f_{i}(\Delta^{\alpha\beta},\theta,\phi)|\vec{p}|$, for $i=1\ldots
N$, and $1\leq N \leq 8.$ $f_{i}(\Delta^{\alpha\beta},\theta,\phi)$
is a real variable and is independent of the momentum magnitude
$|\vec{p}|$ because the photon is massless in the Lagrangian of
Eq.~(\ref{Lagrangian}). So the physically free photon velocity is
\begin{equation}\label{groupSpeed}
c_{\gamma
i}\equiv\frac{dE}{d|\vec{p}|}=f_{i}(\Delta^{\alpha\beta},\theta,\phi),\quad
\textrm{for } i=1\ldots N,\quad 1\leq N \leq 8, \label{ci}
\end{equation}
which means: (i) The free photon propagates in the space with at most
8 group velocities. (ii) For each mode, the light speed $c_{\gamma
i}$ might be azimuthal dependent and not a constant. As we have
known, the light spreads with different group velocities for
different directions in the anisotropic media in optics. In analogy,
we may view the space-time as a kind of media intuitively. However,
there are essential differences between the optical case and the
photon case here, because all the consequences of the $N$ modes and
the light speed anisotropy are results from the Lorentz invariance
violation or the space-time anisotropy suggested by the new
framework.

We need to clarify some essential points concerning the light
speeds, i.e., $c_\gamma$ in our work and the conventional light
speed constant $c$. $c_\gamma$ is determined by the Maxwell
equations or the propagator in QED, and represents the real
propagation speed of the photon or the Electromagnetic wave freely
propagating in the space-time, whereas $c$ is related with the
Lorentz group and the space-time metric, and serves as a constant.
These two speeds are regarded as the same thing generally, but we
should make clear that they are two different concepts. In the
natural units, $c=1$. When we write it explicitly in any unit
system, the metric is $g_{\alpha
\beta}=\mathrm{diag}(1,-1/c^2,-1/c^2,-1/c^2)$, so $g^{\alpha
\beta}=\mathrm{diag}(1,-c^2,-c^2,-c^2)$. We see that the light speed
$c$ is related with the unit definitions of the time and the space.
And an element $R^{\alpha\beta}$ of the Lorentz group is defined as
the one which satisfies $g_{\beta \nu}R^{\alpha
\beta}R^{\mu\nu}=g^{\alpha \mu}$ where $c$ is invariant, so we call
$c$ Lorentz invariant constant. In this article, we do not consider
the light speed $c$ in our derivation, i.e., we set $c=1$ in the
natural units. Instead, the light speed implied in our arguments is
actually the propagating velocity $c_\gamma$ of the photon or the
Electromagnetic wave, and generally $c_\gamma\neq c$ here.

We show now that our theory suggests the light speed anisotropy with
respect to the azimuthal angle in an ``absolute'' reference frame.
To understand the azimuthal distribution of the GRAAL data, let us
consider two simple 
forms of $\Delta^{\alpha \beta}$.
One is
\begin{equation}\label{DeltaStrain}
\Delta^{\alpha \beta}=\xi m^{\alpha} n^{\beta},
\end{equation}
where $m$ and $n$ are two unit vectors in the space-time and $\xi$
measures the magnitude of LV. When $n$ and $m$ are parallel,
$\Delta^{\alpha \beta}$ of Eq.~(\ref{DeltaStrain}) represents that
there exists a strain along the direction $n$ in the space-time.
This case of the LVM can help us to check whether there is a
preferred direction $n$ in the space-time. When $n$ and $m$ are
orthogonal, Eq.~(\ref{DeltaStrain}) represents a shear in the plane
spanned by the two vectors $m$ and $n$~\cite{Hestenes86}. Another
useful parameterization for $\Delta^{\alpha \beta}$ is
\begin{equation}\label{DeltaTrans}
\Delta^{\alpha \beta}=\lambda k^{\alpha}, \quad k^2=\pm1,
\end{equation}
which represents a translation along a direction $k$ and $\lambda$
measures the magnitude of LV too. Timelike unit vectors can be
parameterized as $(\cosh{\zeta}$,
$\sinh{\zeta}\sin{\theta}\cos{\phi}$,
$\sinh{\zeta}\sin{\theta}\sin{\phi}$, $\sinh{\zeta}\cos{\theta})$,
while spacelike ones as $(\sinh{\zeta}$,
$\cosh{\zeta}\sin{\theta}\cos{\phi}$,
$\cosh{\zeta}\sin{\theta}\sin{\phi}$, $\cosh{\zeta}\cos{\theta})$,
where $\zeta$, $\theta$ and $\phi$ are  three variables to
parameterize the unit vectors.

Now, we assume that there is a preferred direction $n=m$ for the
space-time. For the sake of generality, we can take this direction
as the $x$-axis, i.e., $\zeta=0$, $\theta=\pi/2$ and $\phi=0$. So
Eq.~(\ref{DeltaStrain}) reads $\Delta^{\alpha
\beta}=\mathrm{diag}(0,\xi,0,0)$, which is substituted into
Eq.~(\ref{det}) and then we can obtain all the two physical
solutions for the light speed $c_{\gamma i}$.
$c_{\gamma1}=\sqrt{1-(2\xi-\xi^2)\sin^2{\theta}\cos^2{\phi}},$
$c_{\gamma2}=\sqrt{1-2\xi\sin^2{\theta}\cos^2{\phi}}.$ Neglecting
the  higher powers of $\xi$, we can get $\delta c_{\gamma
a}/c_\gamma\equiv|c_{\gamma \mathrm{max}}-c_{\gamma
\mathrm{min}}|/c_\gamma\propto |\xi|$ and $\delta c_{\gamma
m}/c_\gamma\equiv|c_{\gamma1}-c_{\gamma2}|/c_\gamma\propto \xi^2.$
$\delta c_{\gamma a}$ and $\delta c_{\gamma m}$ represent the
differences resulting from the angular distribution and the mode
differences respectively. So we find two interesting results: (i) The
light speed difference between two modes is proportional to the
square of the element of the LVM, i.e.  $\delta c_{\gamma
m}/c_\gamma \propto \xi^2$. (ii) For each mode, the light speed may
be direction dependent, and this anisotropy is linearly proportional
to the element of the LVM, i.e. $\delta c_{\gamma a}/c_\gamma\propto
|\xi|$. When $\xi=0$, the light speed $c_{\gamma i}$ is equal to the
constant $c=1$, and the angle distribution of $c_{\gamma}$ is a
sphere of radius 1 in the space. But it is direction dependent now
for  $\xi \ne 0$. Along the direction $n$, the light speed decreases
($\xi>0$) or increases ($\xi<0$), and the distribution for
$c_{\gamma}$ is not spherical any more.

For $\Delta^{\alpha \beta}$, the sum of the two above cases reads
\begin{equation}\label{LVM_GRAAL}
\Delta^{\alpha \beta}=\left(
                       \begin{array}{cccc}
                         0 & 0 & 0 & 0 \\
                         0 & \xi & 0 & 0 \\
                         \lambda & \lambda & \lambda & \lambda \\
                         0 & 0 & 0 & 0 \\
                       \end{array}
                     \right),
\end{equation}
which means $n=m=(0,1,0,0)$ in Eq.~(\ref{DeltaStrain}) and
$k=(0,0,1,0)$ in Eq.~(\ref{DeltaTrans}). This $\Delta^{\alpha
\beta}$ represents that there is a preferred direction $n=(0,1,0,0)$
and a translation along $k=(0,0,1,0)$ for the space-time, meaning the space-time is
not isotropic now. The equation (\ref{det}) is so complicated that
we can hardly solve all the eight analytical solutions for the light
speed in Eq.~(\ref{groupSpeed}). We get two solutions. One of them
is physical and its explicit form is also lengthy
\begin{equation}
c_\gamma=\frac{[\sin{\theta}(\sin{\phi}+\cos{\phi})+\cos{\theta}]\lambda^2-\sin{\theta}\sin{\phi}\lambda
-\sqrt{h}}{-1+\lambda^2}\label{aniso3}\\
\end{equation}
with
\begin{eqnarray*}
h&=&1+[2\sin^2{\theta}(\cos^2{\phi}-\sin{\phi}\cos{\phi}-1)-2\sin{\theta}\cos{\theta}\sin{\phi}]\lambda\\&&
+[\sin^2{\theta}\sin{\phi}(\sin{\phi}+2\cos{\phi})+2\sin{\theta}\cos{\theta}(\sin{\phi}+\cos{\phi})]\lambda^2\\&&
+(-1+\lambda^2)(2\xi-\xi^2)\sin^2{\theta}\cos^2{\phi}.
\end{eqnarray*}
Finally, Eq. (\ref{xCE0}) becomes
\begin{equation}\label{xCE}
\delta x_{\mathrm{CE}}=-\frac{4A E_0}{m_e}\beta^2\gamma^3(c_{\gamma}-c'),
\end{equation}
where $c'$ is an effective constant and $c$ is close to 1. The light
speed $c_{\gamma}$ is the specific form of Eq.~(\ref{aniso3}).

Up to now, we are ready with all materials we need. So we can
establish the bounds on the violation variables in
Eq.~(\ref{aniso3}) now, and then use Eqs.~(\ref{aniso3}) and
(\ref{xCE}) to interpret the experimental data.

The GRAAL data of the years 1998-2002 were reported in
Ref.~\cite{Gurzadyan05}, and the bound $3\times10^{-12}$ was given
for the Lorentz violation parameters. The data of the period
1998-2005 were reported in Ref.~\cite{Gurzadyan07}, from which
Fig.~\ref{fig1} in our paper is extracted, as representing the CE
data of the years 1998-2005. In paper Ref.~\cite{Gurzadyan10}, the
setup of the experiment had been improved, then the data of the year
2008 were obtained from the improved experiment, and the paper also
provided the conservative bound $1\times10^{-14}$ on the Lorentz
violation parameters. Fig.~\ref{fig2} is from
Ref.~\cite{Gurzadyan10}, representing the data of the experiments in
2008. The recent publication~\cite{Bocquet10} provided detailed
discussions on a sample of all the 2008 data, and reported the upper
bound of the order of $1.6\times10^{-14}$ (95\% C.L.) on the Lorentz
violation parameters, related to the light speed anisotropy of the
Lorentz violation model therein. The upper limits in
Refs.~\cite{Gurzadyan10,Bocquet10} are almost two orders stronger in
power than that of the previous work in
Refs.~\cite{Gurzadyan05,Gurzadyan07} with the help of experimental
improvements.

Based on Eq.~(\ref{xCE}) here and Eq.~(7) in Ref.~\cite{Bocquet10},
we can write the competitive upper limits on the violation
parameters $\lambda$ and $\xi$ of our Lorentz violation model too
\begin{equation}\label{constraint}
\left|\sin{\phi}(1-\sin{\phi}-\cos{\phi})\lambda-(\cos^2{\phi})\xi\right|<2.4\times10^{-14}
\quad (95\% \textrm{ C.L.}),
\end{equation}
in the case of $\theta=\pi/2$. For comparison, the upper constraints
on the element $\xi$ of the photon LVM are given in
Tab.~\ref{tabel_xi2} from some other experimental results on the
light speed anisotropy.
\begin{table}
  \centering
  \caption{Constraints on the element $\xi$ of the photon LVM from some light
  speed anisotropy experiments.}\label{tabel_xi2}
\begin{tabular}{c|cc}
  \hline \hline
  Experiment & $\delta c_{\gamma a}/c_\gamma$, $|\xi|$  &  \\
  \hline
  Refs.~\cite{Gurzadyan05,Gurzadyan07}  & $3\times 10^{-12}$& one-way \\
  Ref.~\cite{Gurzadyan10} & $1.0 \times 10^{-14}$ & one-way\\
  Refs.~\cite{Riis88,Bay89} &  $3\times 10^{-9}$ & one-way \\
  Ref.~\cite{Krisher90}  &  $3.5\times 10^{-7}$ & one-way\\
  Refs.~\cite{Herrmann09,Herrmann05} (cf. \cite{Stanwix06,Stanwix05})& $3\times10^{-17}$ &two-way\\
  \hline \hline
\end{tabular}
\end{table}

The two figures, Fig.~\ref{fig1} and Fig.~\ref{fig2}, show the
consistency between the two periods of the experiments in the years
1998-2005 and the year 2008. Due to the potentially systematic
errors from the experiments for the CE data in Fig.~\ref{fig1}, the
error bars may be underestimated, and the data of 2008 in
Fig.~\ref{fig2} may have the same problem. So the evidential
regularity of oscillation shown in Figs.~\ref{fig1} and \ref{fig2}
still needs to be confirmed by more precise experiments in the
future. Nevertheless, it is still enlightening for us to obtain the
best-fit values of the Lorentz violation coefficients too, besides
the upper constraints on the violation parameters $\lambda$ and
$\xi$ we have gotten.

We fit Eq.~(\ref{xCE}) with the experimental results presented in
Figs.~\ref{fig1} and \ref{fig2}, in which the solid curves represent
the GRAAL data from Refs.~\cite{Gurzadyan07,Gurzadyan10}. The dashed
curves are the calculated results of Eq.~(\ref{xCE}), and they are
obtained to fit the experimental curves. The bright-color solid
curves are the calculated results averaged over $90$ degrees to fit
the experimental curves too, and they are completely allowed within
error bars by the GRAAL data. In Fig.~\ref{fig1}, the best-fit
parameters are: $\xi=-2.89\times10^{-13}$,
$\lambda=6.53\times10^{-14}$, and in Fig.~\ref{fig2},
$\xi=-3.64\times10^{-13}$, $\lambda=8.24\times10^{-14}$. We also
find that the best-fit occurs when $\xi\simeq -4\lambda$. In this
article, we can take the average and get $\xi=-3\times10^{-13}$ and
$\lambda=7\times10^{-14}$ . So the best-fit LVM for photons can be
approximated by
\begin{displaymath}
\Delta^{\alpha \beta}=\left(
                       \begin{array}{cccc}
                         0 & 0 & 0 & 0 \\
                         0 & -3\times10^{-13} & 0 & 0 \\
                         7\times10^{-14} & 7\times10^{-14} & 7\times10^{-14} & 7\times10^{-14} \\
                         0 & 0 & 0 & 0 \\
                       \end{array}
                     \right)
\end{displaymath}
and $\delta c_{\gamma a}/c_\gamma\simeq10^{-14}$-$10^{-13}$. The
magnitudes of these LVM elements are consistent with the constraints
of various experiments~\cite{photon2}. Finally, we find that if the
tiny anomaly exits for the light speed isotropy in GRAAL
experiments, the corresponding best-fit violation parameters
$\lambda$ and $\xi$ are still allowed by the conservative upper
bounds Eq.~(\ref{constraint}).

Under the strong upper constraint at the level $10^{-14}$, we should
be serious to treat even tiny anomaly for the light speed
anisotropy. The available GRAAL results of the azimuthal
distribution have been explained by the light speed anisotropy
suggested by the new theory of the Lorentz invariance violation. The
best-fit violation parameters $\xi$ and $\lambda$ are compatible
with the upper limits shown in Eq.~(\ref{constraint}), so more
experiments are needed to determine whether the evidential light
speed anisotropy exists or not.

Finally, we present a summary of our findings:

\noindent (i) Even if for a given type of particles (photons here),
different Lorentz violation models have their own Lorentz
parameters, but the same GRAAL experiment provides the upper limits
of the same order for the different Lorentz parameters from
different models. We have seen that the GRAAL data in
Refs.~\cite{Gurzadyan10,Bocquet10} give the constraints at the level
$10^{-14}$  for the Lorentz parameters of the model therein, and
same data also give same constraints for $\lambda$ and $\xi$ of our
model too.

\noindent (ii) The regularity implied by the GRAAL data in
Figs.~\ref{fig1} and \ref{fig2} is stimulating for us to suggest an
anomaly for the light speed anisotropy. In our framework, the
Lorentz invariance violation or the space-time anisotropy for the
photon is the source of the light speed anisotropy. At the same
time, we also get the upper limits on the Lorentz violation
parameters of the new model. It is a surprise that our new model
calculations can reproduce the possible azimuthal oscillation in the
reported GRAAL data of the years 1998-2008 in an elegant manner. As
the oscillation of the light speed anisotropy extracted from
Figs.~\ref{fig1} and \ref{fig2} is allowed by the corresponding
upper constraints also, it is unclear whether the anomaly for the
light speed anisotropy really exists or not. Therefore more
experiments are highly demanded to clarify the situation.

\noindent (iii) This work not only manifests the elegant application
of the new theory to fit the experimental results, but also suggests
new chances to test the theoretical predictions from the new
framework and to constraint the newly introduced Lorentz invariance
violation matrix by future experiments.




{\bf Acknowledgements}: This work is partially supported by National
Natural Science Foundation of China (Grants Nos.~11021092,
10975003, 11035003, and 11120101004) and by the Research
Fund for the Doctoral Program of Higher Education of China.

\bibliographystyle{unsrt}

\end{document}